\documentclass[aps,preprint]{revtex4}
\usepackage{epsfig}
\usepackage{amsmath}
\usepackage{epsfig}

\begin{document}
\title
{\bf Anti-symmetric square well and barrier potential between two rigid walls}    
\author{Zafar Ahmed$^1$, Shashin Pavaskar$^2$, Indresh Yadav$^3$, Tanayveer Bhatia$^4$}
\email{1: zahmed@barc.gov.in, 2: spshashin3@gmail.com, 3: iykumarindresh288@gmail.com, 4: tanayveer1@gmail.com}
\affiliation{$^1$Nuclear Physics Division, Bhabha Atomic Research Centre,Mumbai 400 085, India.\\ $^2$National Institute of Technology, Surathkal, Mangalore, 575025, India.\\
$^3$Solid State Physics Division, Bhabha Atomic Research Centre,Mumbai 400 085, India.\\
$^4$Birla Institute of Technoligy and Science, Pillani, 333031, India.}
\date{\today}
\begin{abstract} 
We study an anti-symmetric (square) well and barrier potential of depth/height $(V_0)$  placed between two rigid walls. Unlike  the usual double-well, here the closely lying sub-barrier doublets need not be the lowest ones in  the spectrum. When $V_0$ admits certain  calculable values, $E=0$ or $E=V_0$ or both could become energy eigenvalues of the special eigenstate which emerge only if one  seeks a linear solution of Schr{\"o}dinger equation in the appropriate regions.
\\ \\
PACS: 03.65-Ge
\end{abstract}
\maketitle
The fact that $\psi(x)=Bx+C$ [1,2] is the explicit zero energy $(E=0)$ solution of  one dimensional Schr{\"o}dinger equation 
\begin{equation}
\frac{d^2 \psi(x)}{dx^2}+[k^2-\frac{2m}{\hbar^2} V(x)] \psi(x)=0, \quad  k=\sqrt{\frac{2mE}{\hbar^2}}
\end{equation}
for free particle (zero potential: $V=0$) 
has revealed the missing eigenstates in the special cases of
well known and often discussed potentials [3-8]. In the case of Dirac Delta well between two rigid walls [3,7], for a special strength parameter, one gets single eigenvalue at $E=0$ which has zero curvature ($\frac{d^2\psi} {dx^2}$) in the domain. In other cases
the new eigenstate has  zero curvature in part(s) of the domain and more interestingly  it may be a ground or any excited state of the potential [4,6]. Recently, it has been shown [8] that the usual rectangular (double well or hole) potential could become special for the certain calculable values of the height and depth parameter $|V_0|$. Then the double well can have the barrier-top eigenstate at $(E=V_0>0)$ as a ground or one of the excited states. Similarly, the  potential hole ($V_0<0$) can have $E=0$ as one of the eigenstates.

In this paper, we place an anti-symmetric well and barrier of
width $2b$ and depth/height $V_0$ at $x=0$  between two rigid walls at $x=\pm a$. We express this potential (see Fig. 1)  as
\begin{equation} 
V(x)=\left\{ \begin{array}{lcr}
\infty, & & x\le -a, x\ge a\\
0, & & -a <x<-b, b<x<a\\
-V_0, & & -b\le x \le 0 \\
V_0, & & 0 \le x \le b \\
\end{array}
\right.
\end{equation}
and propose to solve its discrete eigenvalue problem completely.
We consider three separate cases (i): When $E=0$, (ii) $E\ne 0, V_0$, (iii) $E=V_0$. This potential (2) is an interesting variant  of the one-dimensional double-well potential. We will see that unlike the usual double well, here the closely lying sub-barrier doublets of the potentials need not be the lowest levels of the potential. Such a potential is discussed [8] mostly as a perturbed infinitely deep potential well. The cases (i) and (iii)
will give rise to the condition such that for a fixed geometry
the infinitely many special values of $V_0$ would enable the potential to have zero-energy or barrier-top eigenvalues apart from the usual eigenvalues which emerge from the case (ii).
\\ 
{\bf Case (i): Boundstate at}$ {\bf ~ E=0}$\\ \\
For the the Schr{\"o}dinger equation (1) in zero-potential region 
and for zero-energy we seek the solution as
\begin{equation}
\begin{array}{lcr}
\psi_<(x)=A(x+a),& &  x\in [-a,-b]\\ \psi_>(x)=D(x-a),& &  x\in [b,a]
\end{array}
\end{equation}
which are compatible and vanish at their respective boundaries at $x=-a$ and $x=a$.
The solution of Sch{\"o}dinger equation (1) in the well and barrier region can be written as
\begin{equation} 
\psi(x)=\left\{ \begin{array}{lcr}
B \sin qx + C \cos qx, & &  -b \le x \le 0\\
B \sinh qx + C \cosh qx, & &  0 \le x \le b\\ 
\end{array}
\right \{, \quad q=\sqrt{\frac{2mV_0}{\hbar^2}}\\
\end{equation}
Note that these two solutions and their derivative match at $x=0$, consistently.
Matching the wave functions and their derivatives at $x=-b$,
we get
\begin{equation}
\begin{array}{c}
A(a-b)=-B \sin qb+C \cos qb\\
A=qB \cos qb+qC \sin qb 
\end{array}
\end{equation}
Similarly at $x=b$ we get
\begin{equation}
\begin{array}{c}
B \sinh qb + C \cosh qb= D(b-a)\\
qB \cosh qb + qC \sinh qb =D
\end{array}
\end{equation}
Here we introduce  $a-b=d$.
In order to get the quantization condition or eigenvalue formula for $E$, one has to eliminate
$A,B,C$ and $D$ from these equations. One can find the ratio $B/C$ from Eqs. (5,6) and equate them to get
the energy eigenvalue equation. This requires a careful handling of denominators involving discontinuous functions $\tan \theta$ and $\cot \theta$  in various cases. We use  the most general method (see Ref. [8] for the square well potential) to treat Eqs. (5,6)  (and Eqs. 12, 13, 19, 20 in the sequel here)  as linear simultaneous homogeneous equations of $A,B,C,D$ and look for their non-trivial ($A,B,C,D \ne 0$) solutions (see the  Appendix in Ref. [8]). This method in our present case, demands that:
\begin{equation}
\left |\begin{array} {cccc} -\sin qb & 
\cos qb & 0 & d \\  q\cos qb & q \sin qb & 0 & 1\\  \sinh q b & \cosh q b  & d & 0 \\ q \cosh qb & q \sinh qb & -1 & 0 \\
\end{array} \right |=0.
\end{equation}
Upon simplification we find the condition on the potential parameter for the existence of zero-energy eigenstate in the double well potential as:
\begin{equation}
2qd \cos qb+ (1-q^2d^2) ~\sin qb + (1+q^2d^2)~ \cos qb ~ \tanh qb=0
\end{equation} 
Let us call Eq. (8) $f(V_0)=0$.
The co-efficient $B,C,D$ in Eqs. (5,6) can be obtained as
\begin{subequations}
\label{allequations}
\begin{eqnarray}
B=[q^{-1} \cos qb-d \sin qb] A, \label{equationa} \\
C=[d\cos qb+ q^{-1} \sin qb] A \label{equationb} \\
D=(dq)^{-1}[ \sinh qb (qd \sin qb -\cos qb)- \cosh qb (qd \cos qb+ \sin qb)] A.
\end{eqnarray}
\end{subequations} 
\\ 
{\bf Case (ii): Boundstate at} ${\bf E \ne 0, V_0}$\\ \\
This potential is not treated exactly in textbooks. Here, we present the analysis that one would usually do ignoring the presence of $E=0, V_0$ eigenstates. Instead of  solutions (3) we now have for $E \ne 0$.
\begin{equation}
\begin{array}{lcr}
\psi_<(x)=A \sin k(x+a),& & x \in [-a,-b]\\
\psi_>(x)=D \sin k(x-a),& & x \in [b,a]
\end{array}
\end{equation}
For the region $-b<x<b$, we choose interesting combinations of 
linearly independent solutions.
\begin{equation} 
\psi(x)=\left\{ \begin{array}{lcr}
B r\sin px + C \cos px, & &  -b \le x \le 0\\
B p\sinh rx + C \cosh rx, & &  0 \le x \le b\\ 
\end{array}
\right \{, \quad p=\sqrt{\frac{2m(E+V_0)}{\hbar^2}}, r=\sqrt{\frac{2m(V_0-E)}{\hbar^2}}.
\end{equation}
Note that these two solutions are  match at $x=0$ consistently.
We match the solutions and their derivatives at $x= -b$, we get
\begin{equation}
\begin{array}{c}
A \sin k(a-b) = -Br \sin p b + C \cos p b\\
k A \cos k(a-b)=r p B \cos p b +  p C \sin p b
\end{array}
\end{equation}
Similarly, the matching conditions at $x=b$ give
\begin{equation}
\begin{array}{c}
p B \sinh r b + C \cosh r b= D \sin k(b-a)\\
rp B \cosh r b + r C \sinh r b= D k\cos k(b-a)
\end{array}
\end{equation}
Again we demand the consistency of the above four Eqs. (16,17) and their non-trivial solutions for $A,B,C,D$, we get 
\begin{equation}
\left |\begin{array} {cccc} -r\sin pb & 
\cos pb & 0 & \sin kd  \\ rp \cos pb & p \sin p b & 0 & k \cos kd \\ p \sinh rb & \cosh rb & \sin kd & 0 \\ r p \cosh rb &  r \sinh rb & -k \cos k d & 0
\end{array} \right |=0.
\end{equation} 
The condition (18) simplifies to:
\begin{eqnarray}
2r(k^2 \cos^2 kd -p^2 \sin^2 kd) \sin pb + 2kpr \cos pb \sin 2kd  + \\ \nonumber
[k(r^2-p^2) \sin pb \sin 2kd +2 p(k^2 \cos^2 kd+ r^2 \sin^2 kd) \cos pb] \tanh rb=0.
\end{eqnarray}
The co-efficients $B,C,D$ are given as
\begin{subequations}
\label{allequations}
\begin{eqnarray}
B= [(pr)^{-1} k \cos qb \cos kd -r^{-1} \sin qb \sin kd]A, \label{equationa} \\
C=[p^{-1}k \sin qb \cos kd + \cos qb \sin kd]A,  \label{equationb} \\
D=[r^{-1} \sinh rb (k \cos qb \cot kd- p \sin qb) -\\ \nonumber p^{-1}\cosh rb (p \cos qb + k \cot kd \sin qb)]A. 
\end{eqnarray}
\end{subequations}
\\ 
{\bf Case (iii): Boundstate at} ${\bf E =V_0>0}$\\ \\
Here we seek the solution of (1) as
\begin{equation}
\begin{array}{lcr}
\psi_<(x)=A \sin q(x+a),& & x\in[-a,-b]\\
\psi_>(x)=D \sin q(x-a),& & x\in[b,a].
\end{array}
\end{equation}
For the region $-b<x<b$, we have
\begin{equation} 
\psi(x)=\left\{ \begin{array}{lcr}
B \sin sx + C \cos sx, & &  -b \le x \le 0\\
B sx + C, & &  0 \le x \le b\\ 
\end{array}
\right\{, \quad s=\sqrt{\frac{2m(2V_0)}{\hbar^2}}=q\sqrt{2}.\\
\end{equation}
We match the solutions and their derivatives at $x= -b$, we get
\begin{equation}
\begin{array}{c}
A \sin q(a-b) = -B \sin sb + C \cos sb \\
q A \cos q(a-b)= sB \cos sb + sC \sin sb.
\end{array}
\end{equation}
Similarly, the matching conditions at $x=b$ give
\begin{equation}
\begin{array}{c}
sB b + C = D \sin q(b-a)\\ sB = q D \cos q(b-a)
\end{array}
\end{equation}
Once again the consistency condition for non-trivial
solutions of $A,B,C,D$ arising from Eqs. (22,23)
yields
\begin{equation}
\left |\begin{array} {cccc}   
- \sin sb & \cos sb & 0 & \sin qd  \\ s \cos sb & s \sin sb & 0 & q \cos qd \\ sb & 1 & \sin qd & 0  \\ s & 0 & -q \cos qd & 0 \\
\end{array} \right |=0.
\end{equation} 
Upon expansion of this determinant we get\\
\begin{eqnarray}
\sqrt{2} \cos sb~ [qb+qb \cos 2 qd +2\sin 2qd]-\sin sb ~[1-3 \cos 2qd +2qb \sin 2qd]=0.
\end{eqnarray}
Let us call Eq. (22) as $g(V_0)=0$.
The co-efficients $B,C,D$ appearing Eqs. (19,20) are obtained as
\begin{subequations}
\label{allequations}
\begin{eqnarray}
B=[(\sqrt{2})^{-1} \cos sb \cos qd -\sin sb \sin qd] A, \label{equationa} \\
C=[(\sqrt{2})^{-1} \sin sb \cos qd +\cos sb \sin qd] A,
\label{equationb} \\
D=[(\sqrt{2})^{-1} (2bq-\cot qd) \sin sb- (1+bq \cot qd) \cos sb] A.
\end{eqnarray}
\end{subequations}
For all the calculations here we work with $2m=1=\hbar^2$ and fix $a=6,b=2 (d=4)$. In the Table I, we list out the first five energy
eigenvalues for various values of parameter $V_0$ in the potential (2)  (Fig. 1). 

First we assume $V_0=.0001$ (very small) to recover from Eq. (15) the well known eigenvalues of particle in an infinitely deep potential: $E_n=\frac{n^2 \pi^2}{144}$. An ordinary value of 5 has been taken for $V_0$ to present the ordinary scenario wherein 
there are two negative discrete eigenvalues in the well of depth 5 units and three discrete positive energy eigenvalues below the 
top of the barrier.

The first four roots of the equation $g(V_0)=0$ (22) are $0.0655,
0.2981, 0.5816, 1.3322$; these values make the potential (2) special wherein the four special potentials respectively have $n=0,1,2,3$ eigenstates at energy exactly equal to the top of the barrier $E_n=V_0$. These special states are discussed in case (i) and presented in  Fig. 2. The last column in the Table I, having $E_*$ gets an entry only if a higher eigenstate $(n>5)$ exists exactly at the barrier-top of the respective barrier. The portions of $\psi(x)$ in the domain $x\in [0,b]$ are linear: $Bx+C$.

The first four roots of $f(V_0)=0$ (8) are $0.3333, 4.09982, 12.7396, 26.31113$ which make the potential (2) special in the other way wherein $E=0$ becomes $n=0,1,2,3$ eigenstate of the respective potentials. These states are plotted in Fig. 3.   The portions of $\psi(x)$ in the domain $|x|>b$ are linear ($Bx+C$), though this linearity is not visible for $x>b$ in the cases $n=1,2,3$.

We find the roots $f(V_0)=g(V_0)$ and get $V_0=0.0359, 0.3125, 4.0894, 12.7361, 26.3123$. Out of these the last three values are very close to the ones listed in the rows 8-11 of the Table I, we find that besides  the zero-energy eigenstates the states with  $n=1,6,12,17$, respectively are the barrier-top states  (see also Fig. 4). So in these cases the potentials become more special as they possess both the zero-energy and barrier-top states. However, the value $V_0=0.0359$ does not make the potential special in either ways. In any case, the condition 
$f(V_0)=g(V_0)$ is actually  inexpiable, it however gave these interesting cases.

We have also confirmed that both the zero-energy and the barrier-top states are orthogonal to every other listed eigenstate of the same potential in the table I. For two ground states ($n=0$) presented in Figs. 2 and 3, we have calculated the uncertainty product: $U=\Delta x \Delta p$. For the barrier-top state (n=0, in Fig. 2) $U=0.5696$ and for the zero-energy state ($n=0$ in Fig. 3) $U=0.5647$. These values are greater than the well known minimum value of $1/2$ ($\hbar=1$). For other interesting uncertainty products including that of the simplest zero-energy eigenstates can be seen in Ref. [10]

We find that when $V_0$ increases calculations require high precision and accuracy for calculating the levels lying deepest
in the well (see the rows 2,8-10 in Table I). The expressions presented in Eqs. (15,16) are very helpful in this regards. For lesser values of $V_0$ the calculations using the determinant (14) and solution of Eqs. (12,13) by Cramer's method
as discussed in Ref. [8] would suffice. 

To the best of our knowledge the (square) well and barrier potential is discussed only as a perturbed infinitely deep potential. However, here we have presented a detailed analysis (see case ii above)  of this potential. Nevertheless, but for the cases (i) and (iii) discussed here the spectrum of this potential may not be complete. The missing zero-energy and barrier-top states would disturb the nodal pattern of the eigenstates wherein according to the oscillation theorem [11] the eigenstate $\psi_n(x)$ must have $n$ number of nodes (zeros).  

\begin{table}

	\centering
	
		\begin{tabular}{|c||c||c||c||c||c||c||c||c|}
		\hline
		S.N. &$~~~~V_0~~~~$ & $E_0$ &  $E_1$ & $E_2$ & $E_3$ & $E_4$ & $E_5$& $E_*$\\
		\hline
		\hline
		1 &0.0001 &0.0685 & 0.2741 & 0.6169&  1.0966 & 1.7135 & 2.4674  &\\
		\hline
		2& 5 & -3.733845 & -0.4354 & 0.4972  &  0.7227 & 1.9639 & 
		2.3852 &
		\\
		\hline
 		3&0.0655 & $V_0$ &  0.2756 & 0.6162 & 1.0979 &  1.7132 & 2.4678 &\\
		\hline
		4&0.2981&0.0124 &$V_0$ & 0.6057  & 1.1216  & 1.7088 &
		2.4763 & \\
		\hline
	   5 & 0.5816 & -0.1096& 0.3349 & $V_0$ & 1.1795 & 1.6970 & 2.5001 &\\
	   \hline
	  6& 1.3322& -0.5809 & 0.4015 & .5112 & $V_0$ & 1.6639  &
	  2.6041& \\
	  
	   \hline
	  7& 0.3333 & $0$ & 0.3027 & 0.6032 & 1.1275& 1.7077 & 2.4784
	   & \\
	   \hline
	   8&4.0998& -2.909757& $0$ & 0.4865  & 0.8392 & 1.9127 & 2.4882 & $E_6$\\
	   \hline
	  9& 12.7396& -11.1434197  & -6.510498 & 0 & 0.53823 &
	  0.88086 & 2.72435 & $E_{12}$\\
	   \hline

	  10& 26.31113&-24.50234846 &-19.139948039 & -10.484039 &$0$  & 0.560662 & 0.8940711 &$E_{17}$ \\
		\hline
     
           12& 0.3125 & 0 & $V_0$ & 0.6047 &  1.1240 & 1.7084 & 2.4771 &\\    
        \hline 
      
		\end{tabular}
		\caption{First six eigenvalues of the well and barrier potential between two rigid walls  (Fig. 1) when $V_0$ is varied. Here $a=6, b=2 (d=4)$. The last column $E_{*}$ indicates the energy levels $E_6, E_{12}, E_{17}$ which equal their respective $V_0$ value. The negative energy states require higher precision in calculations.}
		\end{table}

\begin{figure}		
\centering
\includegraphics[width=10 cm,height=5 cm]{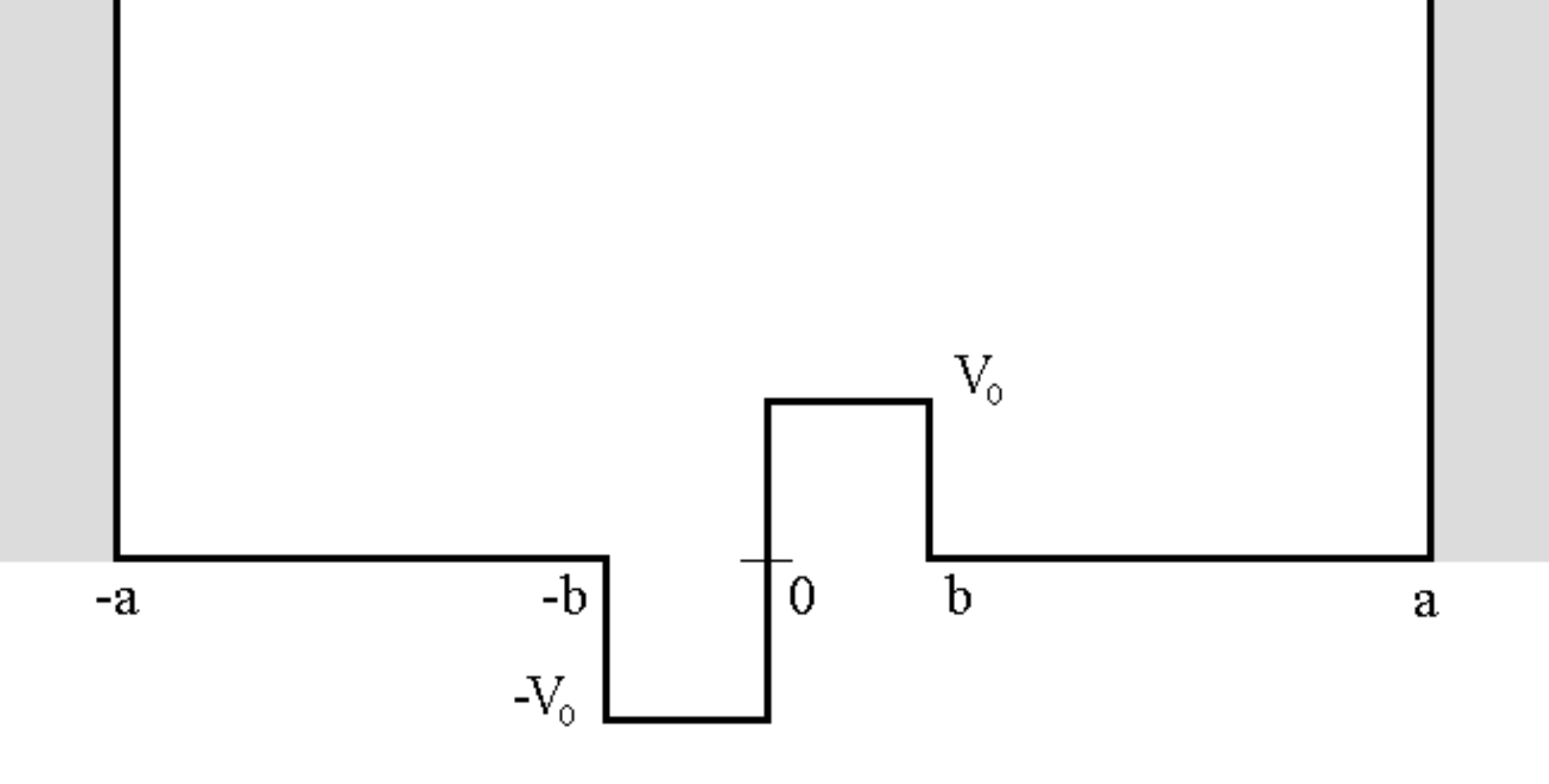}
\caption{Schematic depiction of the square well and barrier
between two rigid walls}	
\end{figure}	
\begin{figure}
\centering
\includegraphics[width=6 cm,height=3.5 cm]{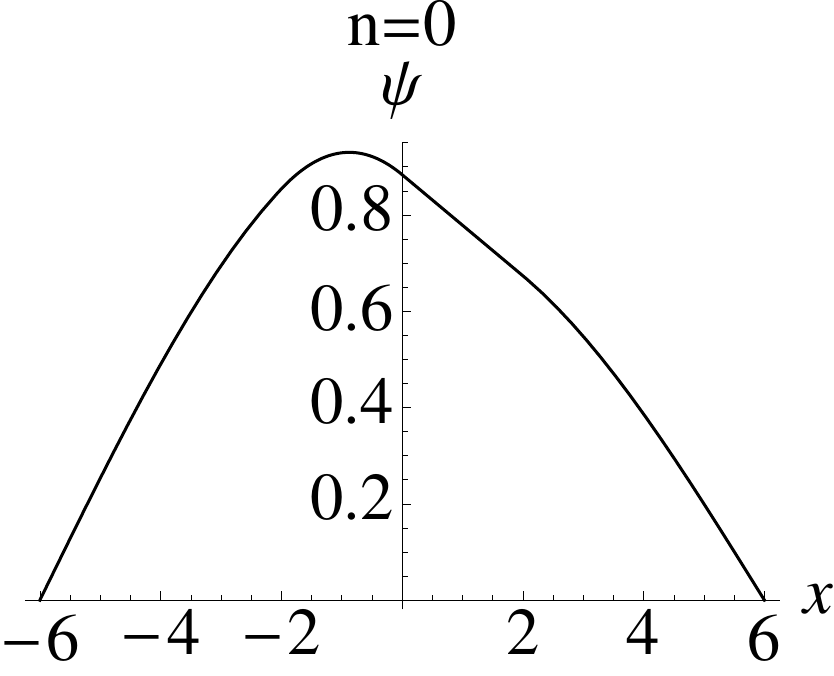}
\hskip .5 cm
\includegraphics[width=6. cm,height=3.5 cm]{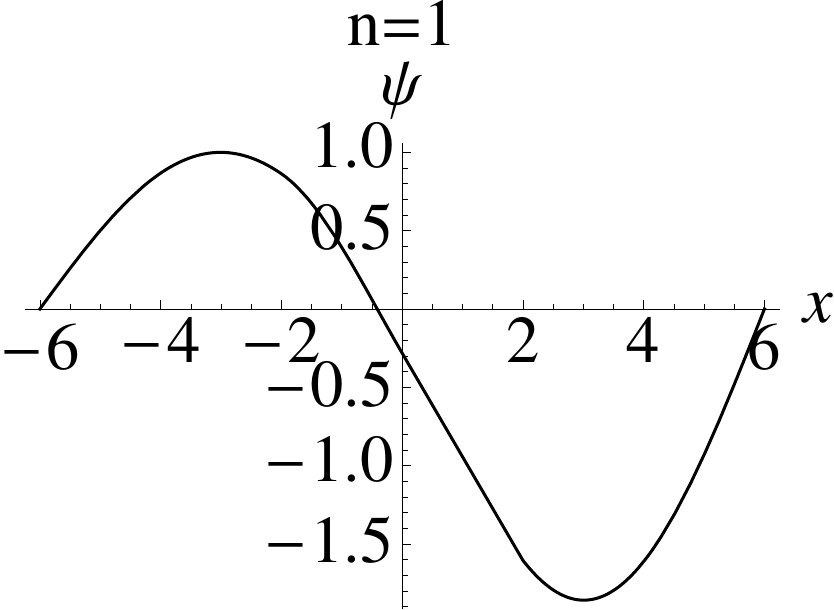}
\hskip .5 cm \\
\includegraphics[width=6.0 cm
,height=3.5 cm]{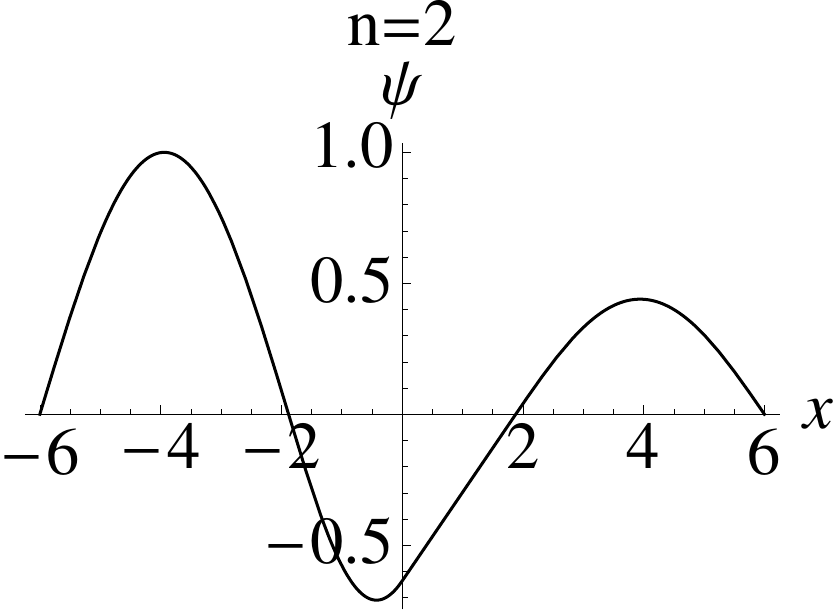}
\hskip .5 cm
\includegraphics[width=6 cm,height=3.5 cm]{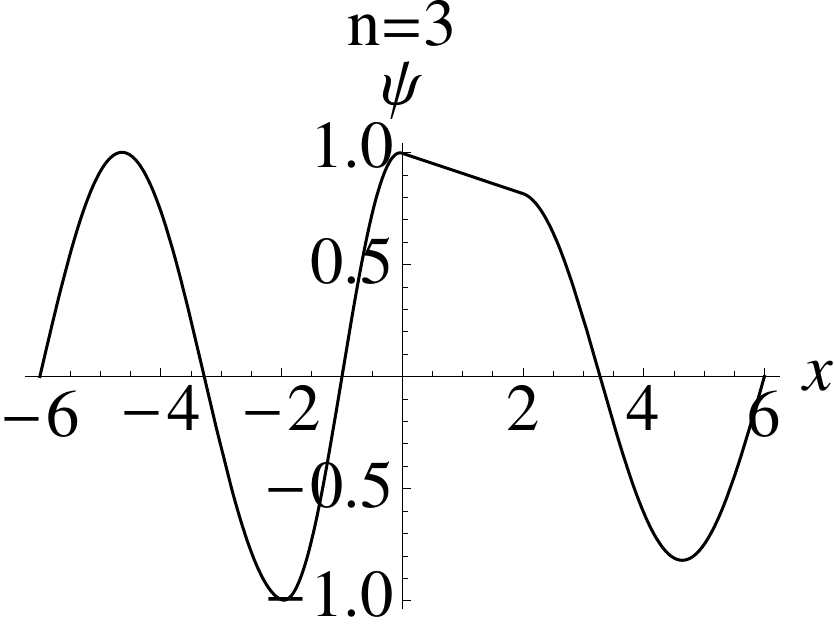}
\caption{The barrier-top un-normalized eigenstates for four values of $V_0$ in the potential (2) (Fig. 1)
See the rows 3-6 of the  Table II. See the linear behaviour 
of the eigenfunctions in the barrier region  $0\le x \le 2$. }
\end{figure}
\begin{figure}
\centering
\includegraphics[width=6 cm,height=3.5 cm]{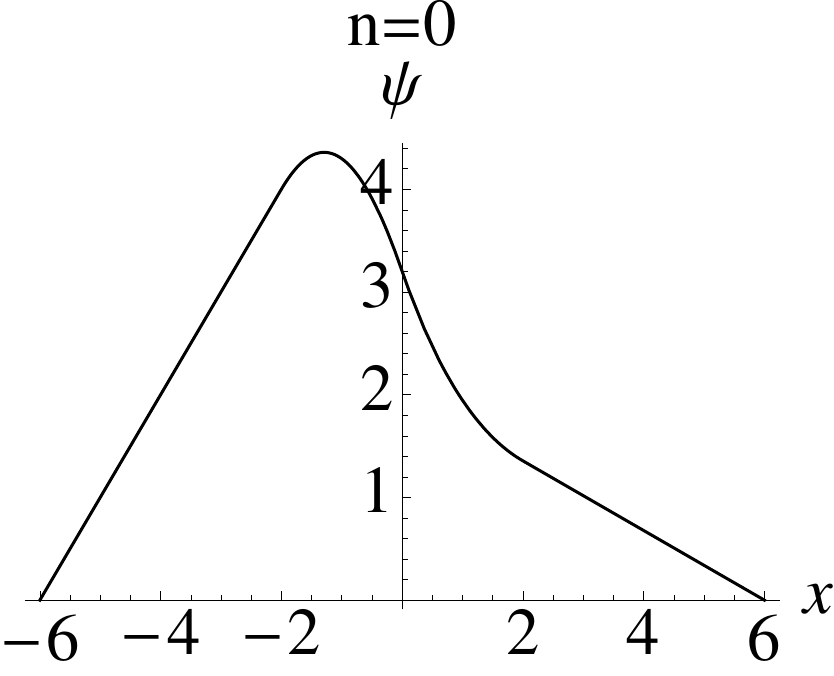}
\hskip .5 cm
\includegraphics[width=6 cm,height=3.5 cm]{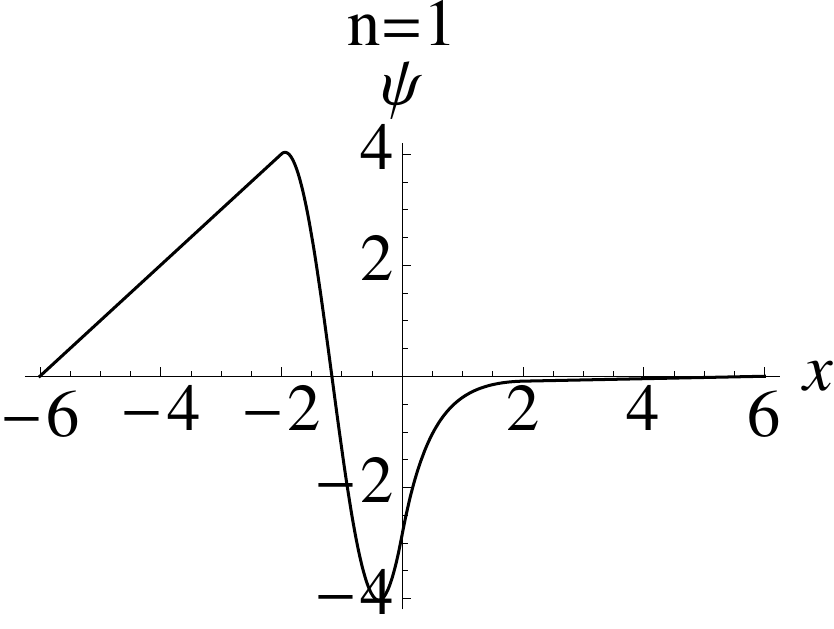}
\hskip .5 cm \\
\includegraphics[width=6 cm,height=3.5 cm]{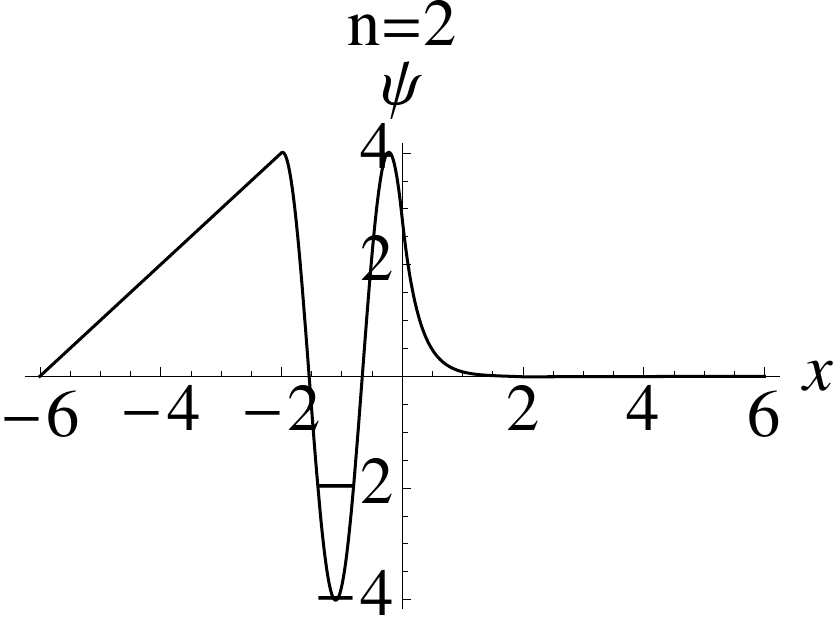}
\hskip .5 cm
\includegraphics[width=6 cm,height=3.5 cm]{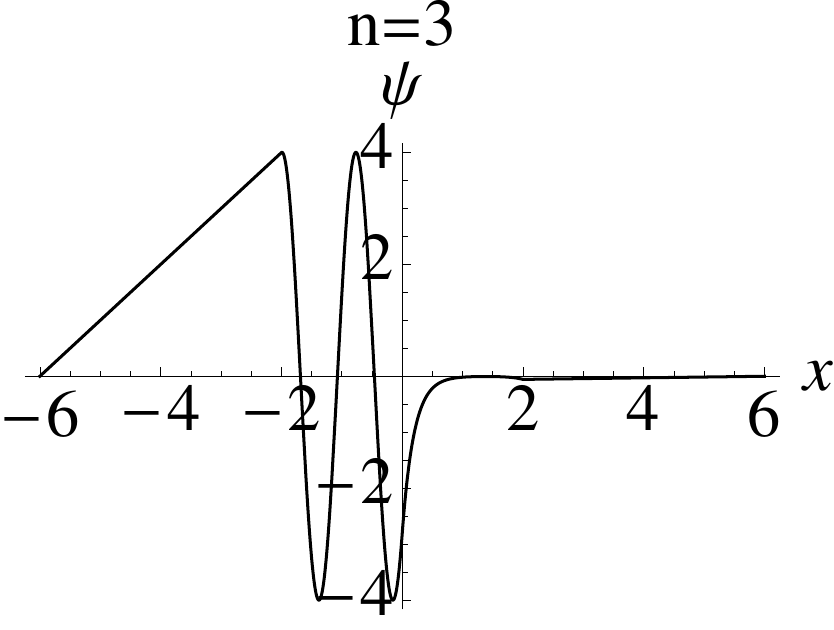}
\caption{The zero-energy un-normalized eigenstates, $\Psi_Z(x)$, of four values of $V_0$ in the  potential 2 (Fig. 1). See rows 7-10 of the Table I. Notice the linear ($Bx+C$) part for $|x|\ge 2$ for $n=0$ state. For other states the linear part is merged with $x$ axis for $x\ge 2$. All these eigenstates are continuous and first-differentiable in the entire domain of $x \in [-6,6]$.}
\end{figure}

\begin{figure}
\centering
\includegraphics[width=6 cm,height=3.5 cm]{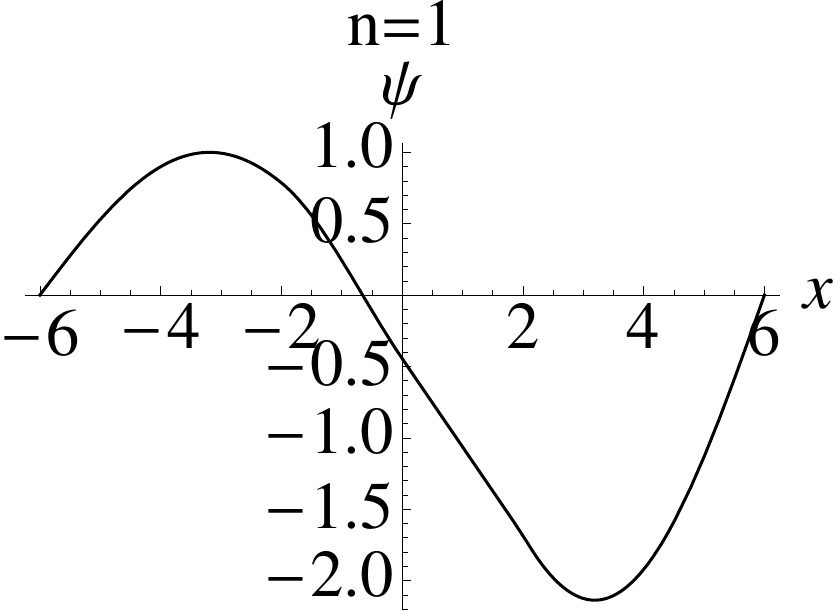}
\hskip .5 cm
\includegraphics[width=6. cm,height=3.5 cm]{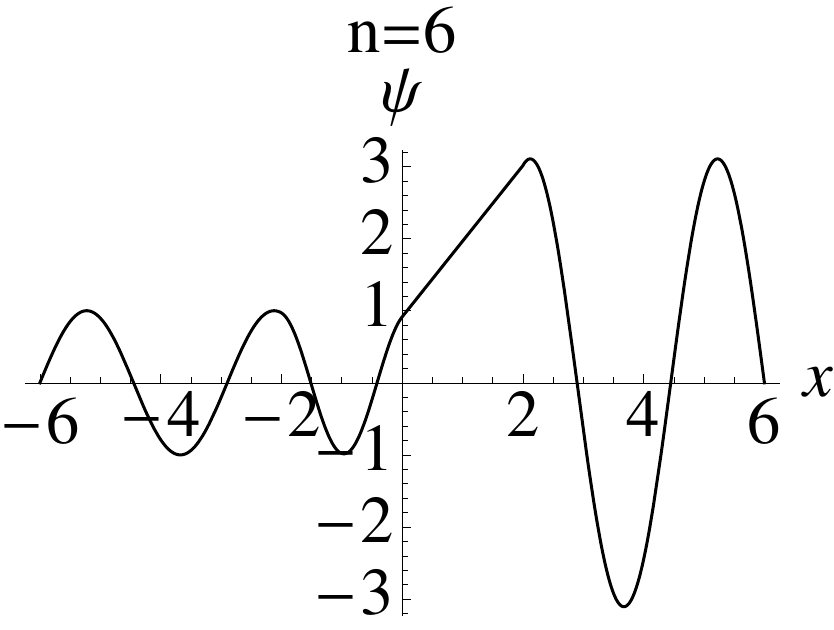}
\hskip .5 cm \\
\includegraphics[width=6.0 cm
,height=3.5 cm]{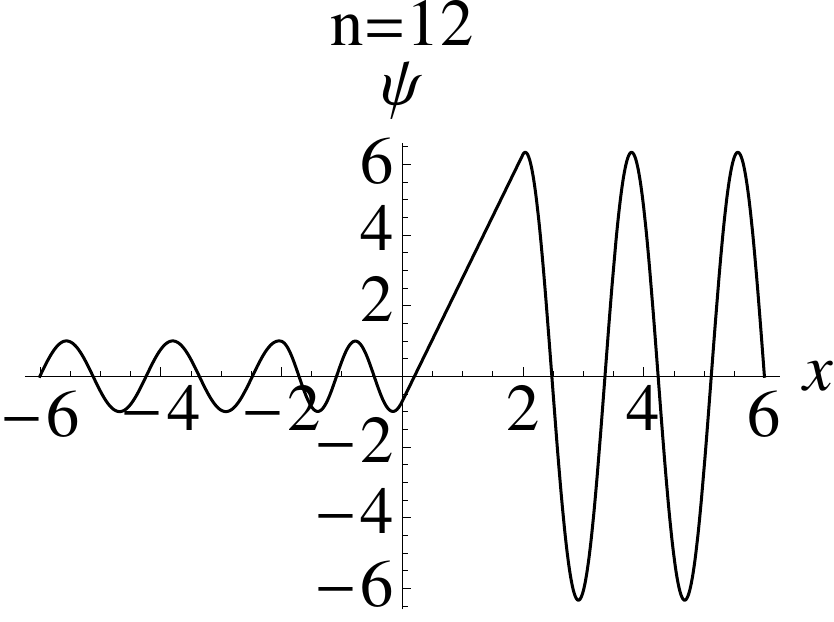}
\hskip .5 cm
\includegraphics[width=6 cm,height=3.5 cm]{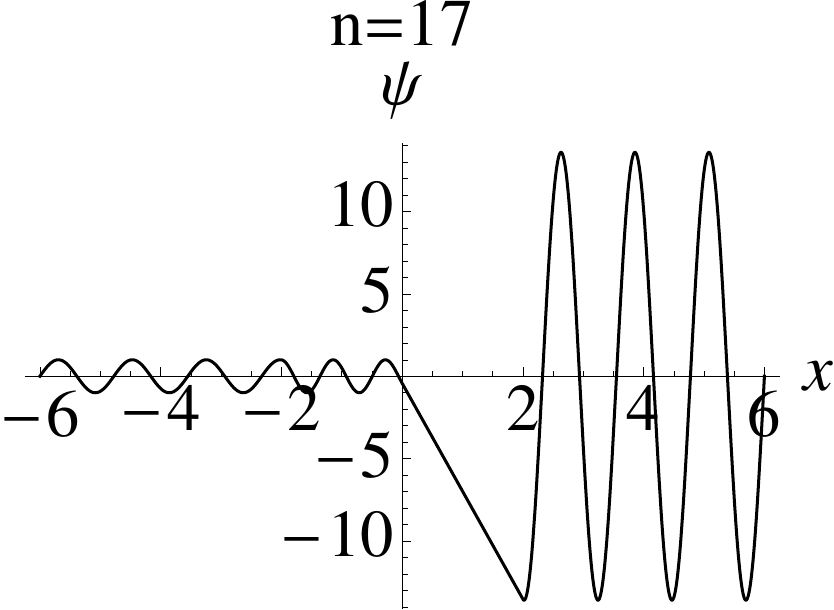}
\caption{The barrier-top un-normalized eigenstates for four values of $V_0$ in the potential (2) (Fig. 1). The potential with $V_0=0.3125, 4.0998, 12.7396, 26.3113$  are even more special as they have both zero-energy $(n=0,1,2,3)$  and barrier-top eigenstates ($n=1,6,12,17$), respectively. See the rows 8-11 in the table I. }
\end{figure}
\end{document}